# Contact-Free Biosignal Acquisition via Capacitive and Ultrasonic Sensors

ROMAN KUSCHE, FABIAN JOHN, MARCO CIMDINS, AND HORST HELLBRÜCK
Center of Excellence CoSA, Luebeck University of Applied Sciences, 23562 Lübeck, Germany

Corresponding author: Roman Kusche (roman.kusche@th-luebeck.de)

This work was supported by the European Regional Development Fund under Grant LPW-E/1.2.1/765.

**ABSTRACT** Contact-free detection of human vital signs like heart rate and respiration rate will improve the patients' comfort and enables long-term monitoring of newborns or bedridden patients. For that, reliable and safe measurement techniques are indispensable. The aim of this work is the development and comparison of novel ultrasonic and capacitive measurement setups, sharing a common hardware platform. Both measurement techniques that are implemented and compared are based on the detection of minor chest wall vibrations in millimeter ranges, due to geometrical thorax changes during respiration and heartbeat activities. After examining the physical measurement conditions and simulating the capacitive sensor, a problem-specific measurement setup is proposed. The system is characterized to be capable of detecting distance changes below 2 $\mu$m via the ultrasonic sensor and below 800 $\mu$m via the capacitive sensor. First subject measurements show that the detection of heart activities is possible under ideal conditions and exclusively with the proposed ultrasonic approach. However, the capacitive sensor works reliably for respiration monitoring, even when the subject is fully-clothed and covered with a blanket. The chosen ultrasonic approach is sensitive regarding minor changes of the reflecting surface and therefore has high uncertainty. In contrast, capacitive respiration detection is very reliable. It is conceivable that improvements in the capacitive sensor circuitry will also enable the detection of heart activities. The proposed ultrasonic approach presents current problems of this technique. In contrast to that, the unusual approach of capacitive sensing demonstrates a high potential regarding vital signs acquisition.

**INDEX TERMS** Capacitive sensor, contact-free, finite element simulation, heart rate, I/Q demodulation, respiration rate, sensor fusion, thorax phantom, ultrasonic sensor, vital signs.

## I. INTRODUCTION

Monitoring of respiration and heart activities are two of the most established biomedical measurement procedures. Current standard techniques focus on functionality rather than on patient comfort. The golden standard methods electrocardiography (ECG) and pneumography guarantee reliable signals, but limit the patient freedom due to the required electrodes, cables, and further equipment [1], [2]. During the past decades, both the sensor technology and the electronic components either analog or digital have been improved significantly. It is therefore conceivable that alternative measurement techniques based on these improvements can combine the required reliability with increased patient comfort. Both heart activity and respiration are of interest in a wide range of medical diagnostics and further applications. In this work, we focus on medical applications, in which the patient is located in a lying position, such as patients in hospital beds or newborns in incubators [3].

To increase patient comfort, the major aim of many recently published methods is the performance of contact-free measurements. One of the most promising approaches is based on video signal processing [4]–[7]. However, the major disadvantages are high costs, computational efforts, and privacy conflicts. Additionally, body parts covered with textiles cannot be used for these visual analyses.

Another widely utilized approach is based on detecting geometrical changes of the chest wall, caused by the heart beat and respiration [8]–[11]. Fig. 1 illustrates this simplified principle. The chest wall raises and lowers by an amplitude of $\Delta d$. These variations include the information of both addressed biomedical signals. Since typical respiration rates are much lower than heart rates, frequency separation is applied [11]. In this example, the geometrical chest wall variations are therefore low-pass filtered (LF) and high-pass filtered (HF).

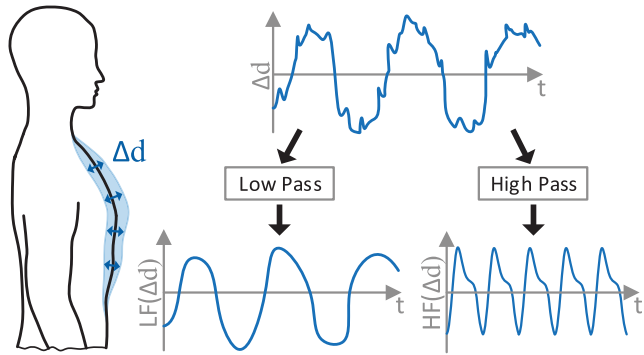

**FIGURE 1.** Principle of biosignal acquisition by analyzing geometrical changes of the chest wall during respiration and heart activities. The low-frequency components are considered to be caused by respiration and the high-frequency components are related to the heart activities.

Related work focuses in particular on techniques like ultrasonic (US) or radio technologies, such as ultra wideband (UWB), microwave sensors, or radio frequency signals in the 2.4 GHz industrial, scientific and medical (ISM) band [9], [10], [12]–[18].

Since the International Agency for Research on Cancer (IARC) of the World Health Organization (WHO) classifies radio frequency electromagnetic fields as possibly carcinogenic to humans, long-term UWB or other radio frequency-based approaches might lead to a low patient acceptance [19], [20]. Regarding the experiments in recent publications, ultrasound-based distance measurements for biosignal acquisitions lead to promising results [9], [10], [21], [22]. However, the number of publications is low and the performed studies do not cover large groups of subjects.

Capacitive sensing is a well-known term and technology in biomedical signal acquisition with a thin layer of isolating material between the human body and the actual sensor element [23]. In this context, the term contact-less is used when the capacitive electrode has no direct contact with the skin [24], [25]. To improve the patients comfort, the electrodes may be integrated into a belt that is attached at the chest [26] or at the abdomen [27], or integrated into a vest [28]. A different capacitive approach is presented in [29]. The sensor is placed below the nose and measures the breath water vapor of the respiration.

In contrast to these approaches, in this work application-specific capacitive proximity sensors measure the distance between a sensor and a body surface. Especially, the significant differences in electrical behavior of human tissue, air, and clothes are a promising advantage of the proposed capacitive proximity sensing approach [30]. This principle is also well-known in short-distance applications like capacitive touch buttons. The approach has also been applied for respiration monitoring, however in a plate capacitor setup with the human between both plates published in [31]. Another approach monitors the heart rate while connected to the fingertip to a capacitive sensor [32]. The approach shown in this work does not require the subject to be between capacitors plates or requires direct contact to a sensor. Instead, we direct the field towards the subject.

The goal of this work is the implementation and comparison of the two most promising non-radio frequency-based principles of contact-less biosignal acquisition with a focus on respiration and heart activities. After a brief introduction to ultrasonic and capacitive sensors including finite element simulations, we present the implementation of a measurement system, that combines both techniques. This system is characterized and finally used to perform first measurements on a human subject. In the result section, we compare the actual behaviors and benefits of the techniques critically. The results of this work are not limited to biomedical applications only. They can also be applied in fitness or automotive applications [33]. In summary, the main contributions proposed in this work are:

- A new capacitive sensor including circuitry for contact-free biosignal acquisition.
- Measurements on a phantom and a subject to compare the characteristics of the capacitive sensor with an ultrasonic sensor.
- Metrological analysis of the robustness of both methods.
- A simple sensor fusion approach to increase the reliability of contact-free respiration monitoring.

## II. MATERIALS AND METHODS
### A. GEOMETRICAL THORAX VARIATIONS

The geometrical changes of the thorax due to respiration or the pumping of the heart vary significantly over time and between subjects [11]. The frequencies of both affect the vibration strength of the observed chest wall. It is also conceivable that the depth of the breathing changes, on the one hand, the low-frequency component of $\Delta d$ (see Fig. 1). On the other hand, mechanical coupling effects between the lungs and the heart within the human body are conceivable to play a substantial role. Regarding the literature, typical chest movements due to respiration vary in a range between $\Delta d \approx 4\ldots12$ mm, whereat the heartbeats lead to vibrations between $\Delta d \approx 0.2\ldots0.5$ mm [11]. Therefore, the goal of the proposed measurements setup is to detect chest movements in this range.

### B. ULTRASONIC DISTANCE MEASUREMENT

The ultrasonic distance measurement is typically based on determining the time of flight of a transmitted pulse between a transmitter (Tx) and a receiver (Rx), as shown in Fig. 2 [34]. With the knowledge of the sound velocity in air, the measured propagation time is used to calculate the distance between the Tx/Rx-combination and the reflecting tissue or other reflecting material. Since the reflecting surface is typically not perfectly even, additional paths (grey, red) exist next to the green drawn direct path. This leads to a superposition of several signal components at the receiver. To distinguish the direct path from the other signals, many implementations transmit short sonic pulses and consider the first received component as the direct path [9], [22]. In applications with multiple active transmitters, a clear allocation of the received signal's origin is difficult. In addition, the generation of very short narrowband transmission pulses limits the maximum local resolution that can be achieved. An approach to circumvent this is employing the frequency separation via a continuously active transmitter. Accepting that the received signal is the superposition of all reflected signal components at the receiver, the processing of a continuous sinusoidal signal Rx(t) is less complex and averaging improves the signal quality [21]. The amplitude $\hat{A}$ as well as the phase shift $\varphi$ of Rx(t) contain information regarding the distance d between the sensor and the reflective surface, as shown in equation 1 [21].

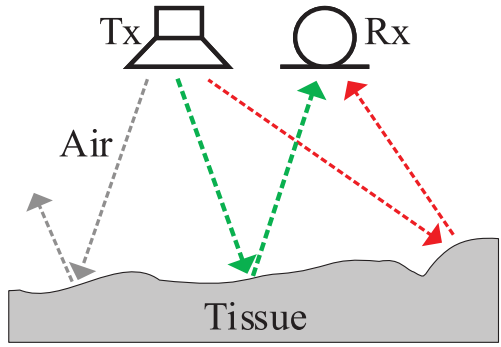

**FIGURE 2.** Simplified principle of ultrasonic distance measurement. The uneven surface of the reflecting material, in this approach human tissue, causes scattering effects. The green dashed arrow represents the desired path to determine the distance between the tissue and the transmitter (Tx)-receiver (Rx) setup.

$$\text{Rx}(t) = \hat{A} \cdot \sin(2\pi f t + \varphi) \quad (1)$$

We assume that the absolute distance d between the subject and the sensor is much larger than the minuscule vibrations on the chest in sub-millimeter ranges. Therefore, the variations of $\hat{A}$, caused by respiration or heart beats, are very small. In addition, $\hat{A}$ might change randomly due to the superposition of the reflected signal components. Taking a typical ultrasonic frequency of $f_{US} = 40$ kHz into account, the corresponding wavelengths $\lambda_{US}$ are close to the vibration dimensions, as estimated in equation 2.

$$\lambda_{US} = \frac{c_{US,\text{Air}}}{f_{US}} \approx \frac{343 \text{ m/s}}{40 \text{ kHz}} \approx 8.6 \text{ mm} \quad (2)$$

Therefore, the phase shift $\varphi$ of the received signal is very sensitive regarding minor distance changes $\Delta d$, independent from the total distance d. Neglecting the missing information of d, which is not of interest in the desired application, this work focuses on this phase shift to extract information regarding chest wall vibrations, caused by respiration and heart activities. In detail, we calculate the relative phase shift from the transmitted and received signal.

### C. CAPACITIVE DISTANCE MEASUREMENT
Capacitive sensing is a wide technical field including many different physical approaches and measurement techniques [35]. Fig. 3 illustrates the principle of a capacitive proximity sensor and the relevant electrical effects [36]. On the right side the sectional view of a circular sensor, consisting of a negative electrode ring and a positive inner circular electrode, is drawn. Additionally, an optional driven shield electrode is shown, which can be used to force the electrical field distribution towards only one direction. The blue lines represent the highly simplified electrical field lines, passing the air with a relative permittivity of $\epsilon_{\text{Air}} \approx 1$ and the subject body with $\epsilon_{\text{Body}}$. Due to the complex composition of different tissue types, the relevant permittivity of the body can vary significantly and depends on frequency. In contrast to water ($\epsilon_{\text{Water}} \approx 81$), the permittivity of the human body can reach up to $\epsilon_{\text{Body}} \approx 10^5$, considering a frequency range of tens to hundreds of Kilohertz in this work [30]. Since the capacitance between positive and negative electrode $C_{\text{Sensor}}$ of the sensor depends on the permittivity ($C_{\text{Sensor}} = f(\epsilon)$) in-between, it is obvious that significantly different permittivities affect the measured capacitance.

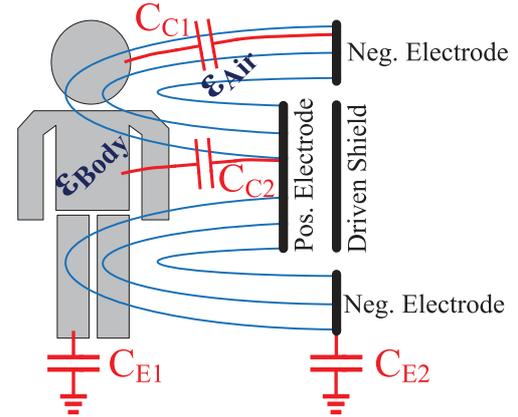

**FIGURE 3.** Principle of capacitive distance measurement. The circular sensor consists of an inner positive electrode and an outer negative electrode ring. An optional driven shield can be used for unidirectional field propagation. This principle is based on the different permittivity values of air $\epsilon_{\text{Air}}$ and the human body $\epsilon_{\text{Body}}$. In addition, the parasitic capacitive coupling of the human body ($C_{E1}$) and the sensor ($C_{E2}$) affects this approach.

The second physical effect is the presence of parasitic capacities. The combinations of conductive sensor electrodes, conductive human tissue and the isolating air in between cause the coupling capacitances $C_{C1}$ and $C_{C2}$, which depend on the distance between subject and sensor and further geometries. Hence, this effect generates information as well. Additional capacitive couplings between the subject and earth ($C_{E1}$) but also between the sensor system and earth ($C_{E2}$) are unavoidable. Since these capacitances depend on complex geometries and many other conditions, the corresponding values vary over time and it is impossible to calculate or predict them. A typical literature value for a rough estimation is $C_E \approx 100$ pF [37]. Assuming realistic geometries of electrodes in cm² ranges with distances of several cm in between, the ideal capacitance of $C_{\text{Sensor}}$ is in very low pF-ranges and therefore significantly lower than these parasitic effects. Since the connected cables and measurement circuitry generate additional unknown capacitances with much higher values, absolute measurements are not useful. Instead, the parasitic effects are assumed to be constant and the transient changes of $C_{Sensor}$ are focused.

To measure the resulting capacitance $C_{Sensor}$, including the described effects, several approaches are possible. One promising method is the determination of the corresponding electrical reactance $X_{C_{Sensor}}$, defined in equation 3.

$$X_{C_{Sensor}} = X_{C_{C2}} + X_{C_{C1}} \parallel (X_{C_{E1}} + X_{C_{E2}}) \qquad (3)$$

A summarized representation of $X_{C_{Sensor}}$, including the frequency dependency, is given in equation 4.

$$X_{C_{Sensor}} = \frac{1}{2\pi f C_{Sensor}} \qquad (4)$$

Technically, $X_{C_{Sensor}}$ can be determined by applying an AC voltage and measuring the occurring electrical current. The usage of a known excitation frequency is advantageous for separating the useful information from signal disturbances, afterwards. Choosing high excitation frequencies of $f \gg 10$ kHz and signal amplitudes in V-ranges lead to comfortable measurable currents in $\mu$A-ranges.

### D. CAPACITIVE SENSOR

For this study, a capacitive sensor has been developed to enable the detection geometrical changes of the thorax. A circular printed circuit board (PCB) with a diameter of 205 mm provides the two required electrodes for capacitive sensing. The diameter of the inner circular positive electrode is 125 mm and is separated from the negative outer electrode ring by a 15 mm isolating gap. To simplify the setup, an active shield is not used.

The behavior of this sensor was evaluated in a finite element analysis using COMSOL Multiphysics. In this simulation, the capacitance of the sensor in air without a human body in the considered area was calculated to be $C_0 = 6.51$ pF. Subsequently, this simulation was repeated considering a human thorax ($\epsilon_{Body} = 81$) below the sensor, as depicted in Fig. 4. For this purpose, an ellipsoid with a length of $l = 0.8$ m, a width of $w = 0.4$ m and a height of $h = 0.2$ m was used. To include the behavior of the parasitic capacitive coupling to ground, this thorax phantom is also connected to GND via a 100 pF-capacitor.

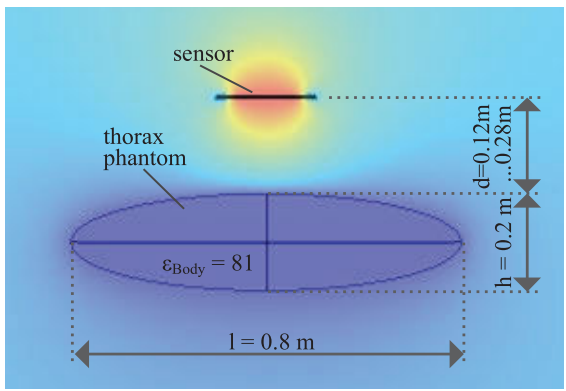

**FIGURE 4.** Finite element simulation model of the capacitive sensor over a thorax phantom.

To determine the influence of the distance $d$ between the sensor and the human thorax phantom, a parametric simulation was performed in the range between $d = 0.12$ m ... 0.28 m, taking the measurement setup in section II-E into account. The simulation results of the sensor capacitance are shown in the left plot (a) of Fig. 5. It can be seen, that the presence of the thorax phantom increases the capacitance by $\approx$ 500 pF. As assumed, this capacitance changes depending on the distance between sensor and phantom. In the right plot (b) the derivation of this function is shown, demonstrating the expected absolute and relative change in capacitance when the thorax is moved. Considering ranges of geometrical thorax movements during respiration and heart activities from section II-A yields relative changes of $\Delta C_{resp.} \approx 400 \ldots 1200$ ppm and $\Delta C_{heart} \approx 20 \ldots 50$ ppm, assuming an average distance of $d = 0.2$ m.

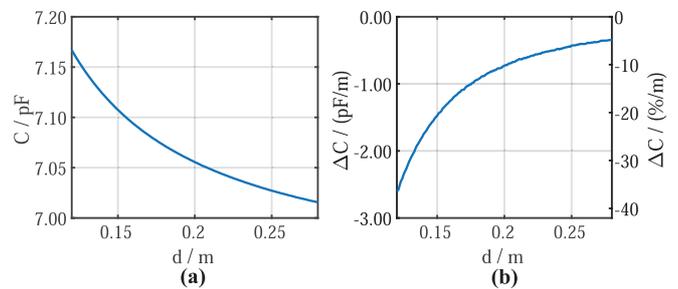

**FIGURE 5.** Results of the finite element simulation of the sensor capacitance depending on the distance between the sensor and the thorax surface. In the left plot (a) the absolute capacitance over a distance range between 0.12 ... 0.28 m is shown. In the right plot (b) the relative changes of the capacitance are presented.

In a second simulation, the influence of the body's permittivity was analyzed. For this purpose, the distance between sensor and thorax was set to $d = 0.2$ m and the permittivity of the body was varied in a range of $\epsilon_{Body} = 10^1 \ldots 10^5$. As it can be seen in Fig. 6, the actual value of $\epsilon_{Body}$ affects the measured capacitance much less that the distance between sensor and thorax. This indicates that the coupling capacitances $C_{C1}$ and $C_{C2}$ in Fig. 3 are decisive for the approach of this work.

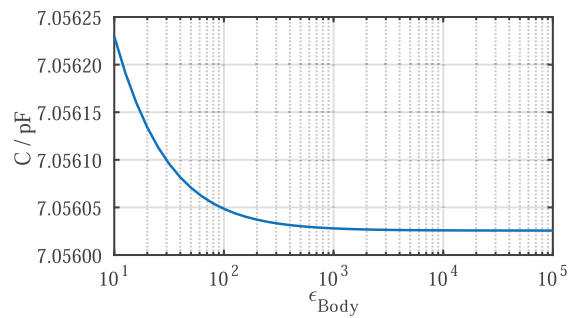

**FIGURE 6.** Results of the finite element simulation of the sensor capacitance depending on $\epsilon_{Body}$.

Since it has to be assumed that the subject does not lie centered under the sensor, the sensitivity regarding a horizontal displacement of $d_{hor.}$ has been simulated. Considering the subject to lie in a hospital bed, a realistic displacement is approximately in the range of $\pm 20$ cm, as shown in the front view in Fig. 7(a). To determine the capacitance changes $\Delta C_{vert.}$ caused by vertical thorax geometry variations, the sweep over $d_{hor.}$ has been performed for d = 19 cm and d = 20 cm. The results of this simulation in Fig. 7(b) show a decrease of the capacitance change and therefore of the sensor sensitivity of approximately 40 % in the considered range of $d_{hor.}$. It can also be seen that the plot is not smooth, which is caused by the limitations of the finite element simulation resolution.

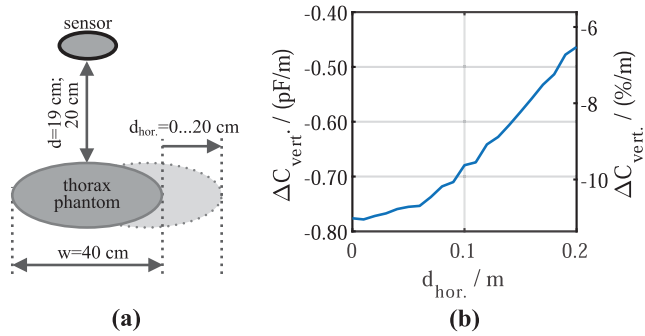

**FIGURE 7.** Front view of geometrical simulation conditions to analyze the effect of a horizontal body displacement (a) and the corresponding simulation results of the sensor sensitivity (b).

Another realistic situation is that the subject lies on the side, causing primarily horizontal expansions of the thorax. In Fig. 8(a) the chosen setup for this simulation is illustrated. The distance between sensor and subject is expected to be constant at d = 20 cm. Only the width changes in the range of w = 20 . . . 21.5 cm. The simulation results in Fig. 8(b) are noisy but show a constant behavior over w, which can be explained on the basis of an ellipsoid's geometry. The changes in the capacitance are $\Delta C \approx 2\%$ and therefore much lower than these of a subject lying on the back (see Fig. 5).

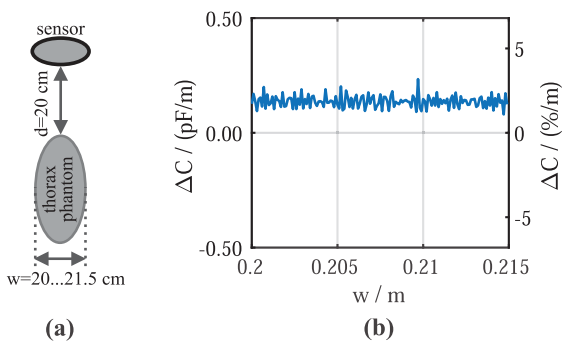

**FIGURE 8.** Front view of geometrical simulation conditions to analyze the capacitance changes when the subject lies on the side (a) and the corresponding simulation results of the sensor sensitivity (b).

In the following sections, this work focuses on situations in which the subject lies on the back, centered under the sensor.

### E. MEASUREMENT SETUP
The proposed measurement setup consists of an ultrasonic sensor, a capacitive sensor, an electronic measurement system, and a host PC, as depicted in Fig. 9.

The ultrasonic sensor contains two identical transceivers (MA40S4R, Murata, Nagaokakyo, JP) with a center frequency of 40 kHz, whereat one works as transmitter Tx and the other one as receiver Rx. Both components are mounted on a breadboard with a distance of 30 mm. As described above, a circular printed circuit board (PCB) provides the two electrodes for capacitive sensing.

The measurement circuitry in the middle of Fig. 9 combines and synchronizes both proposed techniques. It is based on a modified, previously published bioimpedance measurement device which is expected to fulfill the requirements regarding sensitivity [38], [39]. To begin a measurement procedure, the host PC sends a request to the communication interface of the system via a USB connection. This command is forwarded to the microcontrollers $\mu C_1$ and $\mu C_2$ (ATSAM4S16C, Microchip Technology, Chandler, AZ, US). The following colored highlighted sections of the system are similar.

In the upper yellow highlighted section, the internal digital-to-analog converter (DAC) of the $\mu C$ generates a sinusoidal 39 kHz voltage signal with a sampling rate of $f_s = 1$ MSPS and a resolution of 12 bit. To remove the high-frequency components of the synthesized signal, a 4th order reconstruction low-pass filter ($LP_1$) with a cut-off frequency of $f_c = 350$ kHz is implemented via two operational amplifiers (OPAMPs, LMH6628, Texas Instruments, Dallas, TX, US). The following passive 1st order high-pass filter ($HP_1$) with a cut-off frequency of $f_c = 200$ Hz is intended to remove the signal's DC offset. Instead of connecting the sending ultrasonic transceiver directly to this voltage, the sinusoidal signal is converted into a 1 mA AC via a voltage-controlled current source (U/I) [40]. The processing of the received reflected ultrasonic signal begins with a passive first-order high-pass filter ($HP_2$, $f_c = 16$ Hz) to remove DC components. Afterwards, the signal is amplified in a first amplification stage ($OPA_1$, OPA743, Texas Instruments, Dallas, TX, US) by the factor of 20. Two further programmable gain amplifiers ($PGA_1$, $PGA_2$, AD8250, Analog Devices, Norwood, MA, US) are utilized for additional amplification of 20. Between these, the passive high-pass filter ($HP_3$, order N = 1, $f_c = 1$ kHz) removes additional DC components, which might occur due to offset errors of the previous electronic components. To enable sensitive phase shift measurements, an analog IQ-demodulation circuit in combination with complex filtering ($LP_{I1}$, $LP_{Q1}$, N=6, $f_c = 1$ kHz) of the in-phase and quadrature signal components, followed by a high-resolution ADC ($ADC_1$, ADS131E06, 24 bit, $f_s = 1$ kSPS, Texas Instruments, Dallas, TX, US) is implemented. This demodulation part and its characteristics have been published, in [38]. The digitized data is then transmitted to the host PC, where the magnitude and phase of the ultrasonic signal are calculated.

The lower green highlighted section, which corresponds to the capacitive sensor circuit is very similar. Therefore, we briefly explain the differences. As stated before, the choice of a high excitation frequency is advisable to measure the low estimated capacitance values. To avoid influences by the harmonics of the ultrasonic signal processing or interferences from the 50 Hz mains frequency, f = 111 kHz is chosen. After passing $LP_2$ (N=4, $f_c = 350$ kHz) and $HP_4$ (N=1, $f_c = 200$ Hz), the positive electrode of the capacitive sensor is voltage driven by a buffer circuit $OPA_2$, OPA2134, Texas Instruments, Dallas, TX, US). The capacitance changes over time modulate the current through the sensor and the shunt resistor ($R_S = 200$ k$\Omega$) in series. The voltage drop across the shunt resistor, which contains the information about the sensor capacitance, is processed as explained for the ultrasonic sensor. The signal is amplified ($PGA_3$, $PGA_4$) and high-pass filtered ($HP_5$, N=1, $f_c = 1$ kHz) to remove DC offsets and 50 Hz disturbances from the mains. It is then I/Q-demodulated and the low-pass filtered ($LP_{I2}$, $LP_{Q2}$, N=6, $f_c = 1$ kHz) in-phase and quadrature components are digitized with a resolution of 24 bits and a sampling rate of $f_s = 1$ kSPS.

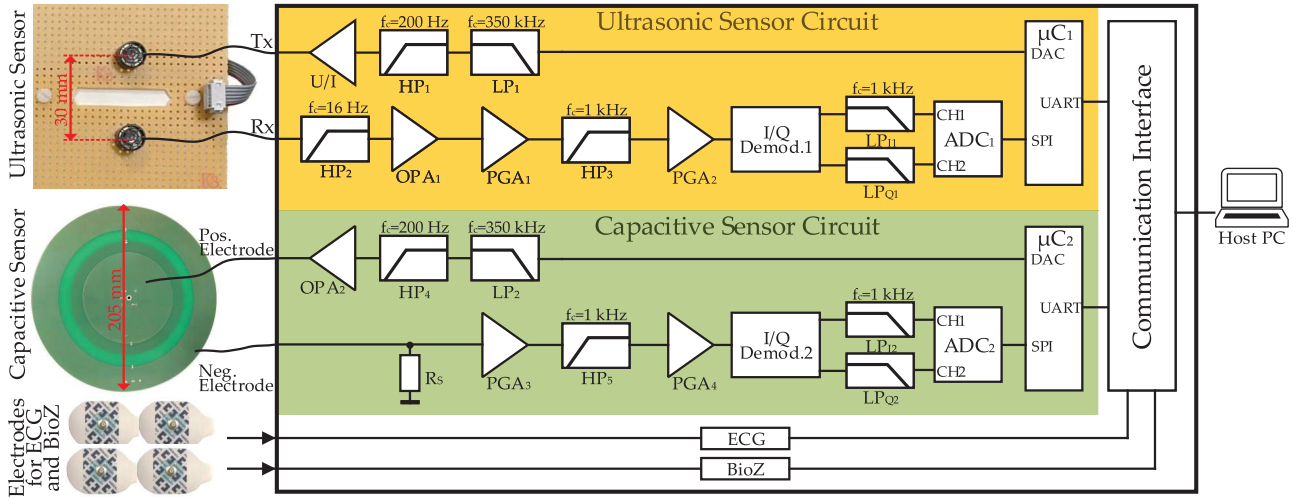

**FIGURE 9.** Measurement setup that combines the ultrasonic and the capacitive sensor concept. On the left side, photographs of both sensors are depicted. The ultrasonic sensor is based on two identical transducers (Rx, Tx), placed on a breadboard with a distance of 30 mm. The capacitive sensor is a circular 2-layers PCB, in which the inner conductive circle represents the positive sensor electrode and the outer ring is used as the negative electrode. The measurement system between the sensors and the host PC generates the required excitation signals, performs analog signal processing and digitizes the data. After transmitting it to the PC, it is further digitally processed via MATLAB.

In addition to these detailed illustrated and described measurement techniques, the original bioimpedance (BioZ) and ECG functionalities of the system are used as reference signals to provide ground truth. Therefore, the additional blocks in Fig. 9 are intended to acquire these signals simultaneously via one common set of electrodes at the subject's chest.

The capability of synchronous acquisition of all signals is used in all subsequent measurements. Plots in a common figures therefore always have the same time reference.

To guarantee the electrical safety of the subject, the measurement system is powered by an external medical power supply (MPU31-102, SINPRO Electronics, Pingtung City, TW) and the USB interface is galvanically isolated. Additionally, the patient auxiliary and leakage currents are below the limits given in the IEC 60601-1. A photograph of a data acquisition PCB is depicted in Fig. 10.

## III. RESULTS AND DISCUSSION
### A. SYSTEM PERFORMANCE
Before performing measurements on a human subject, the characteristic of the sensor setup is analyzed. Both proposed approaches do not analyze absolute measurement values but are based on transient changes. Therefore, a calibration is not necessary. However, the sensitivities of the sensors regarding thorax geometry changes are of interest. For this purpose, a phantom capable of emulating minuscule chest wall vibrations is developed. Fig. 11 illustrates this phantom in combination with both the sensors, mounted on a sensor carrier above.

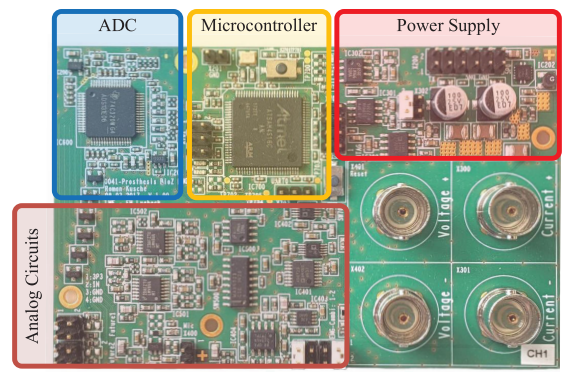

**FIGURE 10.** Photograph of a data acquisition PCB.

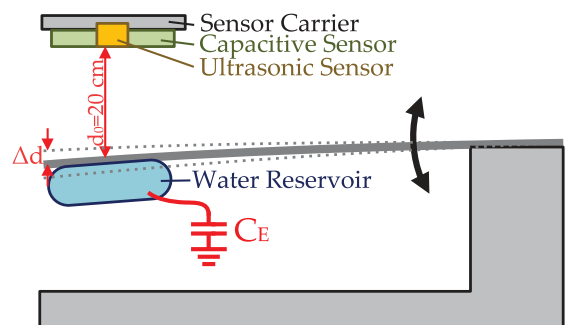

**FIGURE 11.** Experimental setup to characterize the system performance of the ultrasonic and the capacitive sensor implementations. To generate minuscule known distance variations, a plastic plate is stimulated to oscillate. For more realistic conditions, a saltwater reservoir as a simplified human body phantom is mounted under the plate. The capacitive coupling to earth is realized via the capacitor $C_E$.

For generating the desired minuscule changes in the distance d, an oscillating plastic plate is used. To simulate an electrical behavior similar to a human subject, a 3 liters saltwater (sodium chloride, 0.9 %) reservoir is mounted under the plate. This solution is connected via $C_E = 100$ pF to earth potential, representing the described coupling effects of human bodies. Since it is difficult to produce small known oscillations in millimeter ranges and beneath, the plastic plate is excited in such a strong way that the first few oscillation amplitudes can easily be measured via a slow-motion camera. With these determined values, the further exponential decrease of the oscillation is extrapolated. The estimated amplitude values are used in the following plots in Fig. 12 to replace the time axis by oscillation amplitude $\Delta d$. The figure shows the measurement results of both proposed sensors, which have been characterized simultaneously. In the upper plot (a), the measured phase shift of the received ultrasonic signal is shown and in the lower plot (b) the voltage amplitude over the shunt resistor $R_S$ is presented. For better visualization and simpler comparison, both plots use arbitrary units (a.u).

First of all, the noise amplitudes (red) of both sensors were determined by a 10 minutes measurement, corresponding to 600,000 single measurements per sensor, during rest position. Afterward, the plastic plate was stimulated. The blue signals represent the acquired positive signal half-waves from the corresponding sensor. As a meaningful value to describe the measurement uncertainty, the crossing of the signal amplitude with the noise threshold is defined as resolution Q. To increase the reliability of detecting this specific crossing, 20 clearly detected amplitudes (magenta) are used to extrapolate the measured signal amplitudes. With this characterization, the resolution of the ultrasonic sensor has been analyzed to be $Q_{US} \approx 2 \ \mu m$, whereat the resolution of the capacitive sensor is $Q_{Cap} \approx 800 \ \mu m$. Comparing these values with the literature values mentioned in section II-A of this paper, the ultrasonic sensor should be capable of detecting respiration as well as the heartbeat activities [11]. It is even 100 times better than required. In contrast, the sensitivity of the capacitive sensor is not sufficient to reach the required resolution for heartbeat detection. However, respiration monitoring should be possible with it, as well.

### B. SUBJECT MEASUREMENTS

The measurements in this work were approved by the Ethics Committee of the Luebeck University of Applied Sciences. To compare both measurement techniques, additional simultaneous electrocardiogram (ECG) and bioimpedance (BioZ) measurements are performed for providing ground truth.

In contrast to ECG, the measurement of the electrical bioimpedance is an active measurement technique. A small known alternating current with a frequency in the range from tens to hundreds of kilohertz is applied via surface electrodes to the tissue of interest. With two additional electrodes, the occurring voltage drop over the area of interest is detected, which depends on the passive electrical characteristics (impedance) of the tissue in between [41]. Since the conductivity of different kinds of tissue varies, geometrical changes within the body cause also impedance changes. Compared to other tissues, the conductivity of blood is very high. This leads to a decrease of the bioimpedance magnitude, measured over the chest, when the heart ejects the blood. In contrast, filling the lungs with isolating air during inhalation increases the impedance over the thorax [41]. The corresponding techniques are often called impedance cardiography and impedance pneumography, respectively [42], [43].

Since in many applications, the subject does not lie bare-chested under the sensor, the following measurements are additionally executed fully-clothed. In Fig. 13 a photograph of the used measurement setup is shown. The healthy male subject (age: 33; weight: 90 kg) lies on the ground and the contact-free sensors are positioned centered above with a distance of $d \approx 20$ cm. For ECG and bioimpedance acquisition, four hydrogel electrodes (Ag/AgCl, Kendall H92SG, Medtronic, Minneapolis, MN,

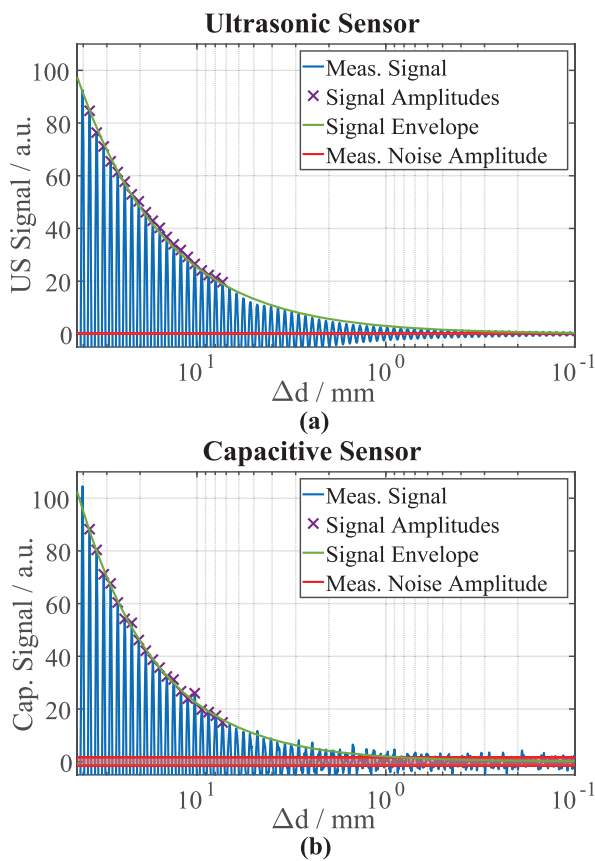

**FIGURE 12.** Measurement results of the system performance of the ultrasonic sensor setup (a) and the capacitive approach (b). For these plots, the time axes are already converted to the corresponding decreasing amplitudes of $\Delta d$. The red areas mark the previously determined noise levels. The crossings of the green envelopes with the red areas are considered to be the resolution limits.

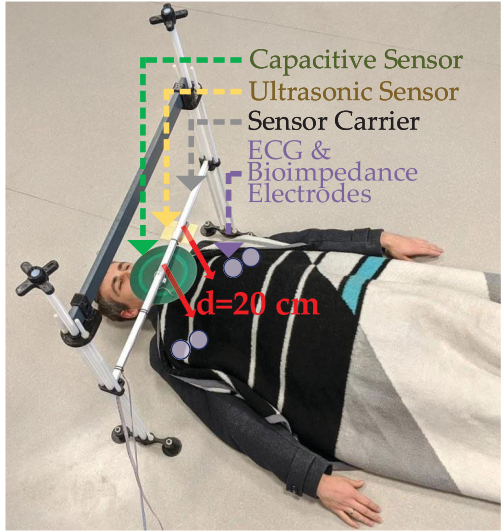

**FIGURE 13.** Measurement setup to acquire respiration and heart activities simultaneously via both proposed approaches. For reference measurements, additionally, the ECG and bioimpedance are measured across the thorax. The air distance between the sensors and the subject is 20 cm. The first measurements are performed bare-chested. Afterward, the procedure is repeated fully-clothed with a T-shirt, pullover, winter jacket, and a blanket, as shown in this photograph.

USA) are positioned on the chest, close to the shoulders, as illustrated.

To produce ideal conditions for heart activity measurements, the subject is bare-chested and asked to breathe in and to hold the breath during the first experiment. The resulting acquired signals are shown in Fig. 14, in which the reference signals ECG and BioZ are filtered to enable a simple visual recognition of the heartbeat rhythm. The signals from the ultrasonic and the capacitive sensor are digitally high-pass (N=2, $f_c = 0.6$ Hz) and low-pass (N=2, $f_c = 6$ Hz) filtered, as well. To avoid influences from the filters' group delays, zero-phase filters are chosen. As shown in several other publications, the ultrasonic signal provides very similar data regarding heart activities [9], [10], [44]. However, even though the proposed implementation of this ultrasonic sensor provides a very high resolution, this particular measurement result was hard to achieve and required several attempts. This problem is reasonable, due to the complex superposition of reflected continuous sound waves. As anticipated, the resolution of the capacitive sensor is insufficient to detect the minuscule chest wall vibrations, caused by the heart beating.

In the next experiment, the bare-chested subject starts breathing after 10 s for 9 respiration cycles. Again, the corresponding reference signals in Fig. 15 are appropriately filtered. Since the total time is 90 s, in the representation of the ECG only the signal peaks can be recognized, whereat the respiration caused bioimpedance changes can be seen. In the ultrasonic signal below, which is high-pass (N=1, $f_c = 0.2$ Hz) and low-pass filtered (N=2, $f_c = 0.5$ Hz), the respiration cycles can hardly be detected. This can be explained by the superposition of all reflected multipath components. The received ultrasound signal is a superposition of all reflected multipath components. Irregular changes in the chest wall geometry while respiration, cause different phase changes in each component and the phase information of the received signal can not be interpreted. In contrast, the capacitive sensor signal is very similar to the BioZ reference. Not only each respiration cycle can be derived, but also the morphology of the signal looks very equal to the BioZ signal.

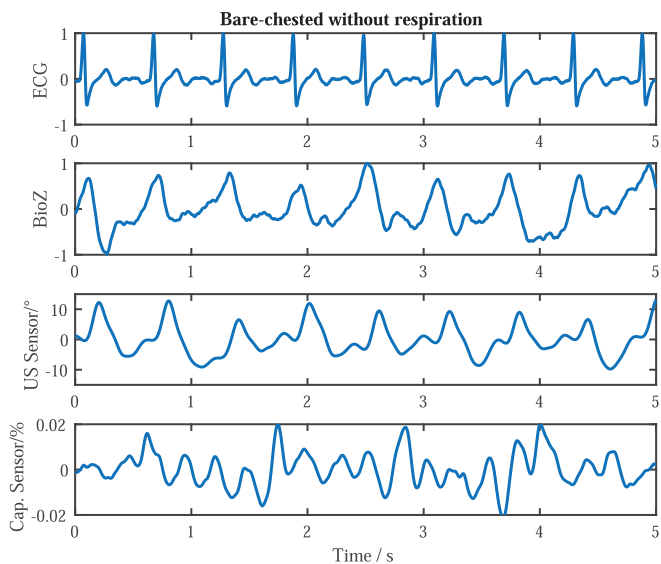

**FIGURE 14.** Measurement results from the bare-chested subject during holding the breath for 5 seconds. In the ECG and the BioZ plots, the expectable heart activity can be seen. The signal from the ultrasonic sensor contains the heartbeat correlated information, as well. However, it cannot be detected in the capacitive sensor signal.

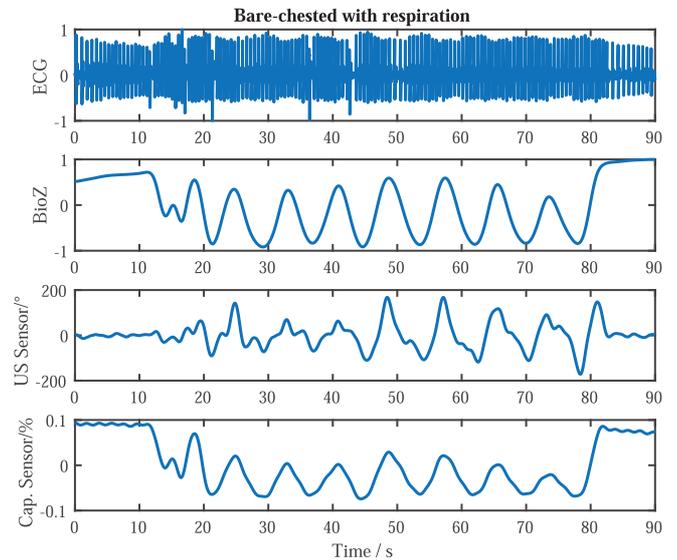

**FIGURE 15.** Measurement result from the bare-chested subject during 9 respiration cycles within 90 seconds. In the BioZ plot, the expectable low-frequency signal can be seen. The signal from the ultrasonic sensor correlates with the respiration, as well. However, strong disturbances in a similar frequency range affect the signal. In contrast, the signal from the capacitive sensor shows a similar morphology like the BioZ signal.

Since the conditions of the subject surface play a decisive role and patients do typically not lie bare-chested in a bed, the previous experiment is repeated under more realistic conditions. For that, the subject wears a T-shirt, pullover and winter jacket. Additionally, he is covered with a blanket, as shown before in Fig. 13.

Similar to the upper experiment, the subject was asked to hold the breath for exclusively measuring the signals caused by heart activities. The corresponding acquired signals, shown for a duration of 5 s in Fig. 16, were filtered in the same way as the signals before in Fig. 14. Even if the subject was covered by several textiles, the signal from the ultrasonic sensor has a strong correlation to the bioimpedance reference signal. However, the morphology of the US signal is different than that of the bioimpedance. The reason could be a stronger influence of scattering effects on the signal, caused by the textiles. As mentioned before, the resolution of the capacitive sensor seems to be insufficient for detecting these minuscule thorax vibrations.

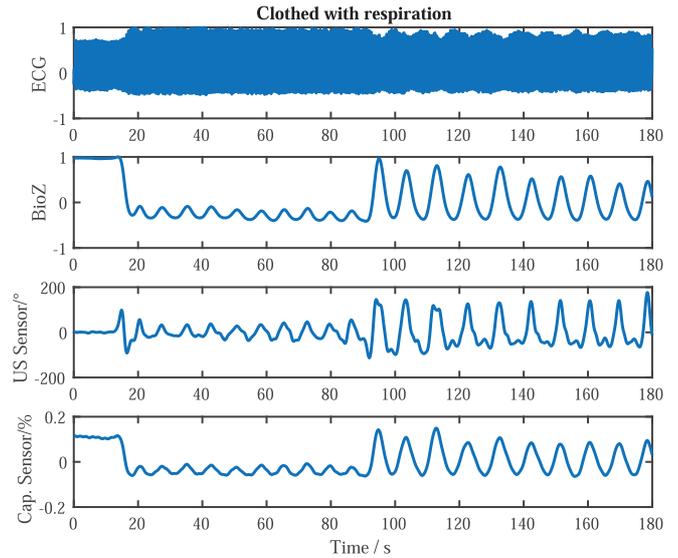

FIGURE 17. Measurement results from the fully-clothed subject during 10 weak and 10 strong respiration cycles within 180 seconds. In the BioZ plot, the expectable low-frequency signal can be seen. The signal from the ultrasonic sensor is affected like in the measurement before. In contrast, the clothes and blanket seem to not affect the capacitive measurement.

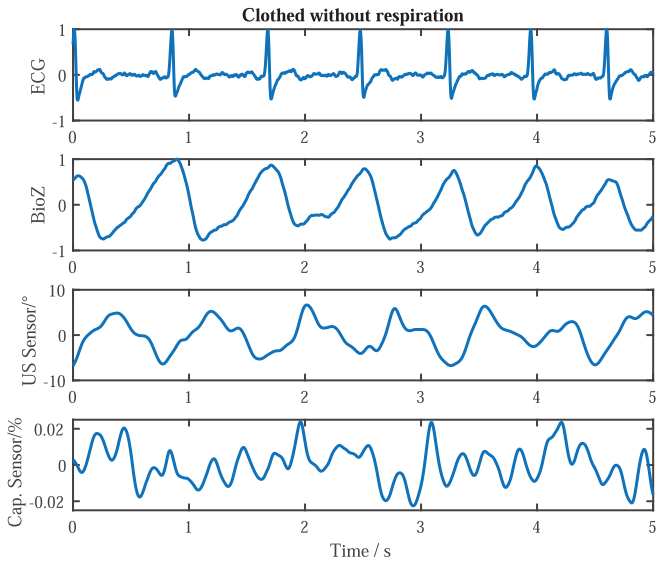

FIGURE 16. Measurement results from the fully-clothed subject during holding the breath for 5 seconds.

In the next experiment, the subject performs 10 weak and 10 strong respiration cycles. The ECG signal in the corresponding results plot in Fig. 17 is depicted for consistency purposes only. Although the distance between the reflecting blanket surface and the chest is $\approx 5$ cm, the respiration cycles are detected via the ultrasound as well as via the capacitive sensor. Since the permittivity of textiles is very similar to that of air ($\epsilon_{\text{Textile}} \approx \epsilon_{\text{Air}}$) and typically electrically isolating, the signal from the capacitive sensor is almost not affected by this covering [45]. The signal morphology correlates to the bioimpedance measurement, which enables further signal analysis. In addition to the respiration rate, the relationship between inhaling and exhaling timings or the strength of breathing could, for example, be of interest in the future.

It has to be considered that the frequency separation between heart activity and respiration signals is challenging. Especially, the previously described fact of significantly stronger chest movements during respiration can affect the simultaneous measurement of the heartbeats. Therefore, it could be useful to focus on higher frequency components of the heartbeat signal, enabling the usage of a higher high-pass filter cut-off frequency $f_c$.

## C. ENVIRONMENTAL INFLUENCES

Obviously, not only the subject under the sensors affects the acquired signals, but also transient changes in the environment can have an influence. In realistic situations this could be a person who is walking close to the sensor. To analyze this condition, a measurement setup as depicted in Fig. 18 is realized. Since a short distance between subject and capacitive sensor causes low coupling capacities, it is assumed that the influences from the environment would be low. Also the expected disturbances in the ultrasonic signals would be low. To generate less ideal conditions, this measurement was performed without the subject and bed. Instead, only the sensors were positioned $h_{\text{Sensors}} = 90$ cm above the ground.

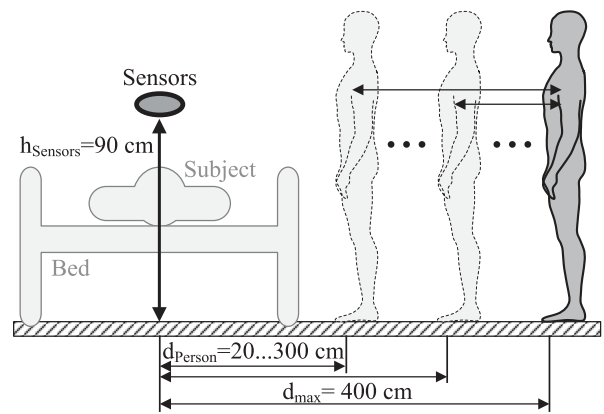

FIGURE 18. Experimental setup to analyze the effect of a walking person close to the sensors. The bed and the subject are only depicted for better visualization.

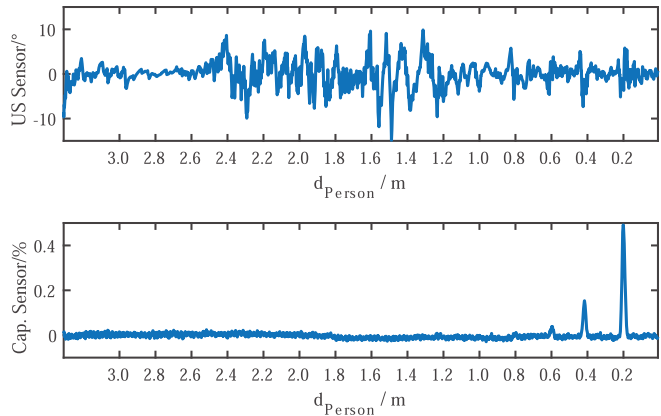

**FIGURE 19.** Measurement results of the ultrasonic and capacitive sensor affected by the walking person. Independent of the actual motions, the disturbances in the ultrasonic sensor signal are in the amplitude range like during heart activity sensing. Critical disturbances in the capacitive sensor signal occur, when the person reaches $d_{Person} \leq 60$ cm.

The person on the right side of Fig. 18 stood $d_{max} = 400$ cm away from the sensor. When the measurement was started, the person began to walk towards the sensor ($T_{Walk1} = 2$ s) and went back to the initial position ($T_{Walk2} = 2$ s). This procedure was repeated for resulting distances between sensors and walking person in the range of $d_{Person} = 20 \ldots 300$ cm with a resolution of 20 cm. The total duration of this measurement was 180 s. Both acquired signals were digitally filtered so that their pass bands cover the frequency range of heart activities and respiration (US: $f_{c,HP} = 0.2$ Hz, $N_{HP} = 1$, $f_{c,LP} = 6$ Hz, $N_{LP} = 2$, Cap.: $f_{c,LP} = 6$ Hz, $N_{LP} = 2$). The resulting signals are plotted in Fig. 19, in which the corresponding values of $d_{Person}$ are noted at the abscissa.

It can be seen that the signal of the ultrasonic sensor varies during the whole measurement between $\pm 10°$, which is also the range of the heart activity measurements shown above. These disturbances seem to be not correlated with $d_{Person}$. This is reasonable since the signal phase is measured and not the magnitude. However, also between the walking periods, when the person stands still, these disturbances are significantly higher than during the measurements in section III-B. One reason could be a lower signal amplitude due to the missing subject, which could lead to a quality loss in the phase analysis. Another origin could be vibrations of the measurement setup in $\mu$m ranges, caused by the steps of the walking person. However, compared to the presented respiration measurements these disturbances are are very low, even if the person walks close to the sensor. These results point out limitations of the ultrasonic approach and question its actual suitability regarding heart beat detection in real environments. Depending on the particular measurement setup, its usage could still be valuable in contact-free respiration monitoring.

In the plot of the capacitive sensor, three signal artifacts can be found which are higher than the noise level: CapSensor($d_{Person} = 0.6$ m) $\approx$ 0.04 % CapSensor($d_{Person} = 0.4$ m) $\approx$ 0.15 % CapSensor($d_{Person} = 0.2$ m) $\approx$ 0.49 % These relative signal amplitudes are in the same range like the measurement results of weak respiration in section III-B. Since it is also conceivable that realistic occurring disturbances are in the low frequency ranges of actual respiration signals, misinterpretations cannot be excluded.

Mechanical barriers like in Fig. 18 the bed itself could prevent the person to get too close to the sensor.

### D. SENSOR FUSION

As shown in the experimental measurements above, changes in the environment lead to signal disturbances and can therefore cause misinterpretations. Due to the different physical measurement techniques, the characteristics of these disturbances differ, as well. This different sensitivity regarding artifacts is utilized for a simple sensor fusion, based on the correlation of both the acquired signals. According to equation 5, the sliding correlation coefficient $C_{Corr}$ of the ultrasonic and the capacitive sensor signals at the sample number $k \geq N$ is calculated under consideration of the last $N = 10^4$ samples, corresponding to 10 s [46]. In this equation, $\mu$ are the mean values and $\sigma$ are the standard deviations of the corresponding signal segments Cap and US.

$$C_{Corr}(k) = \frac{1}{N-1} \sum_{i=k-N+1}^{k} \left( \frac{Cap_i - \mu_{Cap}}{\sigma_{Cap}} \right) \left( \frac{US_i - \mu_{US}}{\sigma_{US}} \right) \quad (5)$$

To evaluate this algorithm, a fully-clothed subject measurement under the same conditions as in section III-B is used. In the total duration of 325 s the time period from $T = 120$ s $\ldots 300$ s corresponds to the previously shown measurement in Fig. 17. The time series can be separated into six periods of different measurement conditions: $T_1$, $T_6$ : Active motions of subject $T_2$, $T_5$ : Strong respiration $T_3$ : Breath holding $T_4$ : Weak respiration In Fig. 20 the corresponding acquired and normalized signals of the bioimpedance reference, the ultrasonic and the capacitive sensor are plotted. Since the disturbances in the capacitive sensor signal are very high, the actual respiration signals are hardly visible in this representation. The bottom plot represents the sliding correlation coefficient. Neglecting the time delay of $C_{Corr}$ caused by the window size of 10 s, it can be seen that during both time periods which are influenced by motion artifacts, the correlation coefficient is $|C_{Corr,T1,T6}| < 0.75$. This low correlation between the sensor signals indicates a low reliability of the data. In contrast to that, the strong respiration during $T_2$ and $T_5$, reliably measured with both sensors, results in a good correlation coefficient of $|C_{Corr,T2,T5}| > 0.87$. This value applies also to the weak respiration during $T_4$. In the orange marked time span, when the subject holds the breath, the correlation coefficient drops below $|C_{Corr,T3}| < 0.6$, due to different noise and distortions of the signals.

Even if this simple signal processing approach is promising, an enhanced validation considering more subjects and different disturbances is advisable. It is also conceivable that alternative sensor fusion algorithms could be useful.

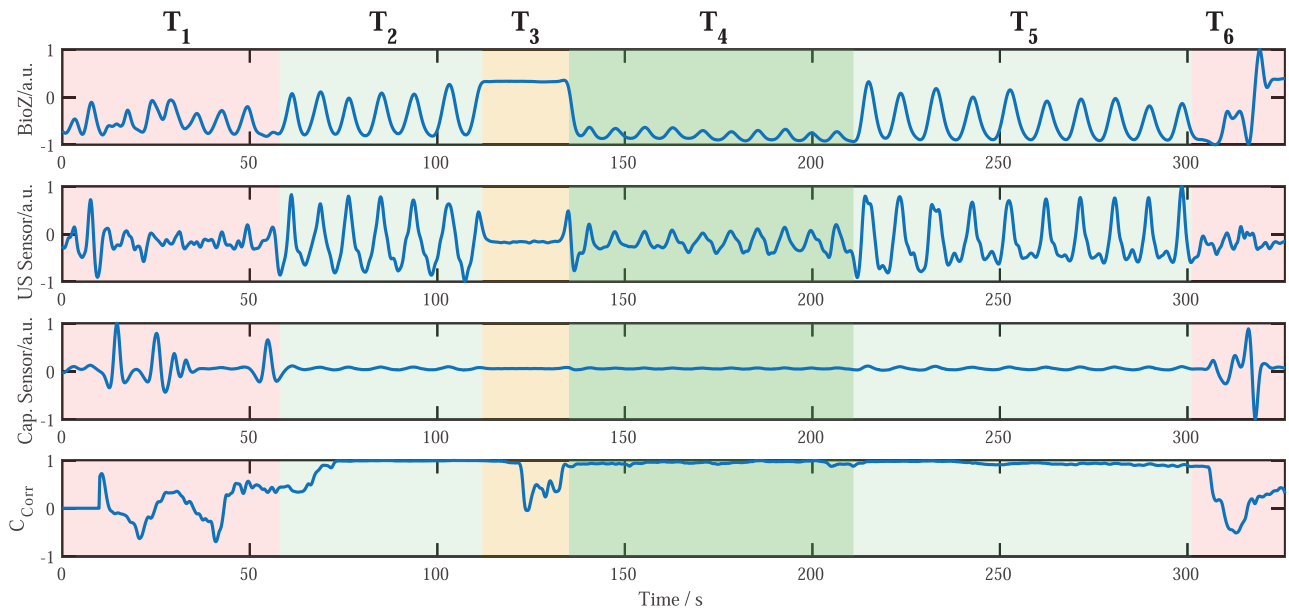

**FIGURE 20.** Results of the sliding correlation coefficient. The time series is separated into the following six periods: $T_1$, $T_6$: Active motions; $T_2$, $T_5$: Strong respiration; $T_3$: Breath holding; $T_4$: Weak respiration.

## IV. CONCLUSION

The results show that both techniques proposed are capable to detect the respiration of a human subject. The high resolution of the ultrasonic sensor enables under ideal measurement conditions additionally the acquisition of the heartbeat.

In contrast to previous studies, the results also indicate a low reliability of the ultrasonic sensor approach, caused by the complex reflection conditions at the body surface. From this point of view, the technique is unpromising to motivate comprehensive subject studies. In the future, beam forming techniques could compensate the described problem by decreasing the observed area and therefore inequalities in the superimposed multi-path components.

In comparison, the results of the capacitive sensor are very positive. As explained in the characterization section, the resolution of this setup is not high enough to detect heart activities. But, the respiration monitoring works reliably and obtains signals which are very similar to that of the bioimpedance reference. It is conceivable that further improvements to the capacitive sensor setup enable the monitoring of heart activities. Additionally, an increase in the distance between subject and sensor is desirable but also challenging as it can be seen in the finite element simulation results. To achieve that, higher electric field strengths and enhanced electronic measurement circuitry are advisable. One approach could be the sigma-delta principle for capacitance measurements which is also implemented in several commercially available microcontrollers [47].

Even if both proposed sensors and their combination via sensor fusion provide useful results, they cannot reach the quality and reliability of conventional techniques. In addition, the morphology of the signals can be influenced by the actual geometrical conditions and is not always equal to signals acquired via conventional methods. In critical medical situations, it is therefore not possible to replace wired measurement systems by the contact-free sensors proposed in this work. However, there are other applications conceivable in which the comfort is more important than the reliability.

The similar frequency range of heart beats and respiration cycles can cause difficulties in separating both signals. Enhanced signal processing algorithms would therefore be useful to increase the reliability.

In summary, the focus of this work was the technical implementation and first evaluation. In the future, further comprehensive subject studies have to be performed. These should also involve subjects with respiratory or cardiac diseases.

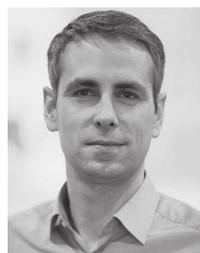

**ROMAN KUSCHE** received the B.Sc. degree in electrical engineering from the Lübeck University of Applied Sciences, Lübeck, Germany, in 2013, and the M.Sc. degree from the Hamburg University of Applied Sciences, Hamburg, Germany, in 2014.

From 2014 to 2019, he was a Research Assistant with the Laboratory of Medical Electronics, Lübeck University of Applied Sciences. Since 2020, he has been a Senior Researcher with the Center of Excellence CoSA, Lübeck. His research interests include the development of novel biomedical measurement methods and the related medical electronic devices.

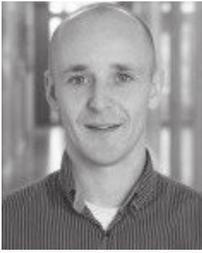

**FABIAN JOHN** received the Dipl.-Ing. degree in electrical engineering from the Hamburg University of Technology, Germany, in 2012. He absolved the certificate program medical physics and engineering at the University of Technology in Kaiserslautern, Germany. He is currently pursuing the Ph.D. degree in electrical engineering with the University of Applied Sciences, Lübeck, Germany.

From 2012 to 2016, he was the Team and Project Leader for test engineering of electronic components for automotive industries at cbb software GmbH, Lübeck. From 2016 to 2019, he was a Development Engineer for production and quality assurance tools for medical laboratory devices at euroimmun, a PerkinElmer company in Lübeck. Since 2019, he has been a Research Associate with the Center of Excellence CoSA, University of Applied Sciences, Lübeck. His research interests include contactless object detection and localization in underwater and medical applications.

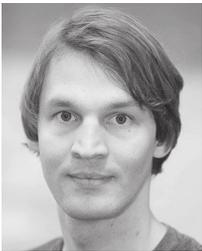

**MARCO CIMDINS** received the B.Sc. degree in electrical engineering from the Lübeck University of Applied Sciences, Lübeck, Germany, the B.S.E.E. degree from the Milwaukee School of Engineering, Milwaukee, USA, in 2015, and the M.Sc. degree in applied information technology from the Lübeck University of Applied Sciences, in 2017, where he is currently pursuing the Ph.D. degree.

Since 2017, he has been a Research Associate with the Center of Excellence CoSA, Lübeck University of Applied Sciences. His research interests include device-free localization, indoor localization, and radio frequency propagation.

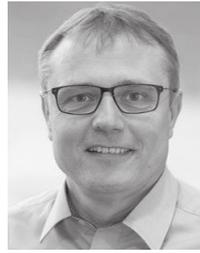

**HORST HELLBRÜCK** received the Diploma degree in electrical engineering from the University of Saarland, in 1994, and the Ph.D. degree in computer science from the Technical University in Braunschweig in the field of ad-hoc networking, in 2004.

He has worked as a Software Engineer with Dräger, Lübeck, from 1994 to 1998, and in International Product Marketing at eupec, Warstein, from 1998 to 2000. He joined the International University in Bruchsal, Germany, in 2000. Before starting his professorship in communication systems and distributed systems at the Lübeck University of Applied Sciences, Germany, in 2008, he held a position as a postdoctoral researcher with the Institute of Telematics, University of Lübeck, Germany. Since December 2013, he has been the Head of the Center of Excellence CoSA and has been appointed as an Adjunct Professor with the Institute of Telematics, University of Lübeck, in 2016. He is currently a Professor with the Lübeck University of Applied Sciences, and an Adjunct Professor with the University of Lübeck. His research interests include modern structures like wireless mobile networks and sensor networks. His special interests include sensor networks in various application fields from medical applications to underwater technologies.

• • •